\documentclass[12pt]{article}
\pdfoutput=1
\usepackage{epsfig}
\usepackage{amsfonts}
\usepackage{amssymb}
\usepackage{xcolor}
\usepackage{mathrsfs}
\usepackage{amsmath}
\usepackage{upgreek}
\usepackage{graphicx}
\usepackage{cite}
\usepackage{calligra}
\usepackage{caption}
\usepackage{subcaption}
\usepackage{cite}
\usepackage{calligra}
\usepackage{hyperref}
\usepackage{tikz-feynman}
\newcommand{\nc}{\newcommand}
\nc{\beq}{\begin{equation}} \nc{\eeq}{\end{equation}}
\nc{\beqa}{\begin{eqnarray}} \nc{\eeqa}{\end{eqnarray}}
\nc{\ba}{\begin{array}} \nc{\ea}{\end{array}}

\def\hc{\ensuremath{\mathrm{h.c.}}}

\newcommand{\tr}{{\text{\,tr}}}
\begin{document}
\begin{center}

{\bf \LARGE Chiral effective potential in $4D,\,\mathcal{N}=4$ SYM theory} \vspace{1.0cm}

{\bf \large I.L. Buchbinder$^{1,2}$, R.M. Iakhibbaev$^{1}$, D.I.
Kazakov$^{1}$, \\[0.3cm] A.I. Mukhaeva$^{1}$,  and  D.M. Tolkachev$^{1,3}$}

\vspace{0.5cm}
{\it $^1$Bogoliubov Laboratory of Theoretical Physics, Joint Institute for Nuclear Research,
  6, Joliot Curie, 141980 Dubna, Russia\\
$^2$ Center of Theoretical Physics, Tomsk State Pedagogical University,
634041 Tomsk, Russia \\
and \\
$^3$Stepanov Institute of Physics,
68, Nezavisimosti Ave., 220072, Minsk, Belarus}
\vspace{0.5cm}

\abstract{We consider $4D$\, $\mathcal{N}=4,\, SU(N)$  super
Yang-Mills theory formulated in terms of $\mathcal{N}=1$ superfields
where the leading low-energy contributions to effective action are
given by chiral effective potential. This effective potential is
calculated in one- and higher-loop approximations. It is shown that
this potential is automatically finite and proportional to the
classical chiral potential. All quantum corrections are found
explicitly and factored into a coefficient at the classical
potential.}
\end{center}

\text{\footnotesize{ $^a$buchbinder@theor.jinr.ru,
$^b$yaxibbaev@jinr.ru, $^c$kazakovd@theor.jinr.ru
$^d$mukhaeva@theor.jinr.ru,}}

\text{\footnotesize{
$^e$dtolkachev@jinr.ru }}

\section{Introduction}

Supersymmetric quantum theory sometimes presents unexpected
surprises. One such surprise is the chiral effective potential
discovered by P. West \cite{West:1990rm}  and later studied some time
ago in works
\cite{JackJW:1990pd,West:1991qt,BKP:1994xq,Buchbinder:1999ui,BuchbinderCvetic}.
It was noted that the chiral effective potential exists only for
massless theories and is always finite\footnote{We take into account that the last integration is always finite if the subdiagrams are made finite} in an arbitrary
supersymmetric theory with chiral-antichiral superfields. The
higher-loop chiral corrections to superfield effective action were
studied in the recent papers
\cite{Buchbinder:2025dgu,Buchbinder:2025jdy}.

The surprise is that at first glance the existence of a chiral effective potential contradicts the
non-renormalization theorem (see e.g. \cite{Superspace:1983nr,WestBook,BK:book})
\footnote{Non-renormalization theorem takes also place in $\mathcal{N}=2$ supersymmetric gauge
theories formulated in terms of harmonic superspace \cite{Buchbinder:1997ib}}. Indeed, according to this theorem,
any contribution to the effective action in the theory formulated in terms of $\mathcal{N}=1$ superfields,
is given only by integral over full superspace, purely chiral and and antichiral contributions are
formally forbidden. However, as pointed out by P. West \cite{West:1990rm}, this theorem does not forbid
the ultraviolet finite space-time nonlocal contributions to one-particle irreducible effective action
in form of integrals over full superspace which can be transformed into local expressions in chiral
subspace due to identity
\begin{equation}
\label{identity}
\int d^8 z ~ u(\Phi) \left(-\frac{D^2}{4 \square}\right) v(\Phi)=\int d^6z ~u(\Phi) v(\Phi),
\end{equation}
where the superspace coordinates are $z=(x,
\theta^\alpha,\theta_{\dot{\alpha}})$ and integration measures are
$d^8z=d^4x d^2\theta d^2\bar{\theta}$ and $d^6z=d^4x d^2\theta$ and
$u$ and $v$ are some functions of chiral superfield $\Phi$. Here
$D^2 = D^{\alpha}D_{\alpha}$, $D_{\alpha}$ is supercovariant
derivative (see e.g. \cite{BK:book}). Thus we see that in principle,
the chiral effective potential is quite compatible with statement of
the non-renormalization theorem. Whenever an identity
(\ref{identity}) arises in the calculation of the effective action,
we can expect contributions to the chiral effective potential to
appear. Also we emphasize that the identity (\ref{identity})
evidently arises only because of structure of the propagators in
massless theories and corresponds to the last integration in the
$L$-loop supergraph when all subgraphs are already renormalized.

In this paper we continue a study, began in the papers
\cite{Buchbinder:2025dgu,Buchbinder:2025jdy}, to construct the
chiral effective potential in the $4D,\, \mathcal{N}=4$ SYM theory.
This theory attracts a lot of attention because it possesses a wide
range of global and local symmetries, it is a finite quantum field
theory, it opens up new possibilities for the development of
non-perturbative approaches in quantum field theory and also it has
multiple relations with string/brane theory. We consider a
formulation of the $4D,\, \mathcal{N}=4$ SYM theory in terms of
$\mathcal{N}=1$ superfields (see, e.g. \cite{Kovacs:1999fx,DHoker:2002nbb}) \footnote{The $4D,\,
\mathcal{N}=4$ SYM theory can also be formulated in terms of
unconstrained $\mathcal{N}=2$ harmonic superfields (see e.g.
\cite{Galperin:2001seg}) where two of four supersymmetries are manifest and another two are hidden.}. In this case the theory is characterized
by well defined chiral-antichiral sector that allows to study the
chiral effective potential.

The main topic of this paper is to study a structure of higher loop
corrections to effective chiral potential in $SU(N)~ 4D,~
\mathcal{N}=4$ SYM theory. The quantum effective action is
constructed in the framework of $\mathcal{N}=1$ superfield
background field method that allows to preserve formally manifest
gauge invariance and manifest $\mathcal{N}=1$ supersymmetry at all
step of loop calculations. This paper generalizes the earlier
results of the work \cite{Buchbinder:2025jdy} where the chiral
contributions to the effective action were calculated in the general
$\mathcal{N}=1$ SYM theory. In this paper, we show that the presence
of extended supersymmetry imposes additional constraints on the
structure of chiral contributions and somewhat simplifies the
calculations and the form of the final results.

The possibility of obtaining chiral loop corrections to the
classical potential was analyzed in Ref.\cite{BKY:1993} where it was
shown that such corrections are possible under the condition $
n_{D^2}+1=n_{\bar{D}^2}$, where $n_D$ is a number of covariant
chiral derivatives in a generic connected 1PI supergraph. Note that
for pure chiral theories, contributions to the chiral effective
potential are only possible beginning with the two-loop level
\cite{BKY:1993}, whereas interaction with the gauge field allows one
to satisfy the above condition even at the one-loop level.

The paper is organized as follows. In Section \ref{sec:nonminN=4} we
describe the  $\mathcal{N}=4$ super Yang-Mills theory formulated in
terms of $\mathcal{N}=1$ superfields and corresponding quantum
effective action in the framework of the superfield background field
method. In Section \ref{sec:loop_corrections} we carry out one - and
two - loop calculations and show that the chiral quantum corrections
indeed exist and are finite. In the last Section \ref{sec:largeN} we
discuss higher loop behaviour of the $\mathcal{N}=4$ supersymmetric
$SU(N)$ Yang-Mills action and its large $N$-color limit. We show
that in all cases the chiral effective potential is equal to
classical chiral potential up to coefficient. All details of quantum
theory are factorized into this coefficient. The results are briefly
discussed in Conclusion \ref{sec:Concl}.

\section{$\mathcal{N}=4$ super-Yang-Mills model}\label{sec:nonminN=4}

\subsection{Effective action and background quantum splitting}
In this section, we consider the $\mathcal{N}=4$  supersymmetric
Yang-Mills theory with the gauge group $SU(N)$ formulated in terms
of $\mathcal{N}=1$ superfields. In such a formulation, the theory
describes the interaction of gauge fields $V$, matter chiral
superfields $\Phi_1$, $\Phi_2$ and $\Phi_3$ and their corresponding
conjugates. Thus the full action of the model can be written as
\begin{equation}
\mathcal{S}_0=\mathcal{S}_c+\mathcal{S}_{GF}+\mathcal{S}_{FP},
\label{eq:action}
\end{equation}
where
\beq
\begin{gathered}
\mathcal{S}_c=\int d^4x ~\mathcal{L}=\tr\int d^8z~ e^{-g V} \bar{\Phi}_i e^{g V}\Phi_i+ \frac{1}{g^2}\tr\int d^6z~\mathcal{W}^2+\\+\frac{i}{3!} g ~\epsilon_{ijk}\tr\int d^6 z~  [\Phi_i, [\Phi_j, \Phi_k]]+\hc,
\label{eq:n4SYM}
\end{gathered}
\eeq
here $W_\alpha=-\frac{1}{8}\bar{D}^2(e^{-2gV}D_\alpha e^{2gV})$. The vector superfield $V$ and the chiral superfield $\Phi$ are in the adjoint representation of $SU(N)$, $g$ is a coupling constant.

 The gauge-fixing term is:
\begin{equation}
\mathcal{S}_{GF}=-\frac{1}{16\xi}\text{tr}\int d^8z~ D^2 V \bar{D}^2 V.
\end{equation}
We work in the Feynman gauge $\xi=1$ since this gauge cancels IR-singularities at the one-loop level \cite{Superspace:1983nr}, this means that the one-loop two-point subgraphs with vector closed lines are identically zero. And finally we have to include the Faddeev-Popov ghost term which is given by
\beq
\mathcal{S}_{FP} =\text{tr}\int d^8z~\left[ \bar{c}'c - c' \bar{c}
+\frac{1}{2}(c'+\bar{c}')[V, c+ \bar{c}] + \cdots \right],
\eeq
Note that $c,c'$  are  (anti)chiral ghost superfields and are given in the adjoint representation of the gauge group.

We also introduce the classical tree-level potential as follows:
\begin{equation}
    W_{tree}=\frac{i}{3!} g ~\epsilon_{ijk}\tr\int d^6 z~  [\Phi_i, [\Phi_j, \Phi_k]].
\end{equation}

Futher calculations are carried out within the framework of the loop expansion of the effective action that is based on the background quantum splitting of the initial fields of the action~\eqref{eq:action}, where the background fields are introduced only for the chiral superfields $\Phi_i$. The gauge superfield $V$ can be only quantum. The corresponding background quantum splitting has the form
\begin{equation}
\Phi_i \rightarrow \Phi_i+\sqrt{\hbar}\phi_i,
\label{chiralsplitting}
\end{equation}
where $\phi_i$ are the quantum superfields and both terms in \eqref{chiralsplitting} transform under the corresponding gauge transformations.

As usual, the quadratic part of the action in the quantum fields $\phi_i$  defines the propagators, and the other part defines the vertices. The corresponding quadratic part of the action~\eqref{eq:action} after background quantum splitting \eqref{chiralsplitting} is written as follows:
\begin{align}
\mathcal{S}^{(2)}&= \tr\int d^8 z ~\left(\bar{\phi} \phi  -2g^2 \bar{\Phi} [v, \phi]+2g^2 \bar{\phi} [v, \Phi] \right) \nonumber\\ \nonumber
    &-g\tr\int d^6z~ \left(\phi_1 \Phi_2 \phi_{3}+\Phi_1 \phi_2 \phi_{3}+\phi_1 \phi_2 \Phi_{3}+ \hc\right) \nonumber \\
    &+\tr\int d^8z~\left(-\frac{1}{2} v\hat{\square}v +cc^\prime+c^\prime c\right), \label{Squadr}
\end{align}
where the quantum gauge superfield $V$ is redefiend as $v$ for uniformity with $\phi_1, \phi_2, \phi_3$. The rest part of the action $\mathcal{S}_c+\mathcal{S}_{GF}+\mathcal{S}_{FP}$  determines interaction vertices of the quantum fields.

\begin{figure}[h]
    \centering
    \includegraphics[width=1\linewidth]{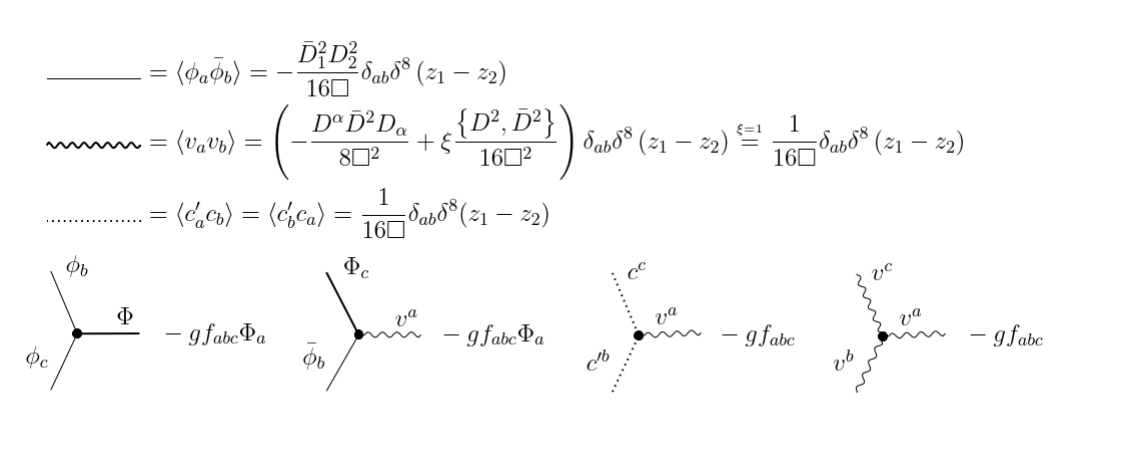}
    \caption{Feynman rules necessary for calculating chiral contributions in the models considered in the main text. External thick lines denote classical background fields, thin lines denote quantum lines. Straight lines are chiral superpropagators, curved line is a vector superpropagator. Black dots denote vertices $g$.}
    \label{fig:Feynrules}
\end{figure}

The details of the deriving Feynman rules from the effective supersymmetric action can be found, e.g., in \cite{BK:book,Superspace:1983nr}. Green functions should not have antichiral external lines. This requirement is equivalent to a restriction on the number of covariant and anticovariant $D$-derivatives mentioned in the Introduction. Keeping this statement in mind, one can obtain Feynman rules for the chiral effective superpotential depicted in Fig. \ref{fig:Feynrules}. Here one should also take into account the corresponding conjugates of the depicted vertices  \cite{BK:book,Grisaru:Improved,Grisaru:SUPERGRAPHITY,WestBook}.
We use the following identites for the $SU(N)$ group generators $[t_a,t_b]=i (T^c)_{ab} t_c$,  $T_{abc}T^{c'}_{ab}=C_A\delta^{c'}_c$, $\text{tr}[t_a t_b]=T_A\delta_{ab}$, where  $C_A=N$.

In the framework of loop expansion, the superfield
effective potential can be written as a series in the number of loops
\beq
\label{loopexp1}
\Gamma[\Phi_i]=\sum_{L=1}^\infty\hbar^L \Gamma^{(L)}[\Phi_i].
\eeq
Therefore, the effective potential $\textbf{W}$ can also be written in the form \eqref{loopexp1}  \cite{BK:book,BKY:1993,Pickering:1996he}, e.g.,
\beq
\label{loopexp}
\textbf{W}[\Phi_i]=\sum_{L=1}^\infty\hbar^L \textbf{W}^{(L)}[\Phi_i].
\eeq

\section{Chiral corrections}\label{sec:loop_corrections}

In this section we calculate chiral corrections to the effective potential~\eqref{loopexp}.

\subsection{One-loop contribution}\label{sec:1loop}

At the one-loop level using the obtained Feynman rules (see Sec.\ref{sec:nonminN=4}) we have the integral which has the following form Fig.\ref{one-loop}.
\begin{figure}[htbp]
 \begin{tikzpicture}[
        scale=1,
        mydot/.style={
            circle,
            fill=white,
            draw,
            outer sep=0pt,
            inner sep=1.5pt
        }
        ]
        \begin{feynman}
            \vertex  at (0,2) (a1){\(p_1+p_2\)};
            \node[below=1cm of a1, mydot] (a2);
            \vertex at (0,1) (a2) ;
            \vertex at (-1, -1) (d); \vertex[below left=0.0cm of d, black, dot] (d){};
            \vertex at (1, -1) (e); \vertex[below left=0.0cm of e, black, dot] (e){};
            \vertex at (-1,-2) (f) {\(p_1\)};
            \vertex at (1, -2) (g) {\(p_2\)};
            \diagram*{
                (a2) --[plain,thick] (a1),
                (d) -- [plain] (a2),
                (a2) -- [plain] (e),
                (e) -- [boson] (d) ,
                (f) -- [plain,thick] (d),
                (g) -- [plain,thick] (e),
            };
            \vertex [left=0.4em of a2] {\(3\)};
            \vertex [left=0.6em of d] {\(2\)};
            \vertex [right=0.6em of e] {\(1\)};
            \node[below=1cm of a1, black, dot] (a2);
        \end{feynman}
         $\hspace{2cm}  \begin{aligned}
   &= \lim_{p_1,p_2 \rightarrow 0} \frac{i}{12}g^3 C_A \int \prod_{l=1}^3 d^8 z_l ~\epsilon_{ijk}[\Phi_{i}(z_1)[\Phi_{j}(z_2), \Phi_{k}(z_3)]]\times\nonumber\\&
  \times \left\{ \frac{1}{\square_2}\delta_{1,2} \frac{D^2_1 \bar{D}_3}{16{\square_1}}\delta_{1,3}\frac{D^2_2}{4}\delta_{2,3}   \frac{1}{\square_2}\right\}=  \frac{\hbar}{(4\pi)^2} \frac{1}{2}g^2 C_A \Upsilon^{(1)} W_{tree},\\
 &   \Upsilon^{(1)}=\int_0^1 d\tau \frac{2 \log(\tau)}{\tau^2-\tau+1},
\end{aligned}
 $
    \end{tikzpicture}
        \caption{Feynman one-loop supergraph contributing to the chiral superpotential.}\label{one-loop}
\end{figure}
Here  $\Upsilon^{(1)}$ is the reduced Davydychev-Usyukina triangle integral \cite{Usyukina:1992jd}. To calculate this integral \eqref{one-loop} first, it is nessesary to integrate over the free delta function and use the standard relation for covariant derivatives $D$, and make use of \eqref{identity}.
After contracting group indices and performing the corresponding $D$-algebra, we make the assumption that the superfields are slowly changing
$\Phi_1(y_1,\theta) \Phi_1(y_2,\theta) \Phi_3(x,\theta)\simeq [\Phi_1 \Phi_2 \Phi_3] (x,\theta),$
where $y_1,y_2,x$ are the 4-coordinates, i.e., all chiral superfields are defined at one point in the full superspace.
Note that in calculations, we first set external momenta to be nonzero, then after calculating the momentum integral, we take the limit of external momenta tending to zero. This leads us to a finite expression which contributes exactly to the chiral effective potential. The described procedure extracts the needed contribution from the effective action and corresponds to the definition of the effective potential.

We take into account possible cyclic permutation of external fields because the gauge propagator can be inserted in every side of the triangle graph under consideration, but as the superfields $\Phi$ and $V$ are in the adjoint representation interacting through the structure constant, the contributions cancel each other.
The finite result~\eqref{one-loop} is the final result for the one-loop contribution to the chiral effective potential and in  agreement with the calculations carried out in Ref. \cite{West:1991qt}.

\subsection{Two-loop contributions}\label{sec:2loop}
\begin{figure}[htbp]
    \begin{tikzpicture}[
        scale=1,
        mydot/.style={
            circle,
            fill=white,
            draw,
            outer sep=0pt,
            inner sep=1.5pt
        }
        ]
        \begin{feynman}
            \vertex  at (0,2) (a1){\(p_1+p_2\)};
            \node[below=1cm of a1, black, dot] (a2);
            \vertex at (0,1) (a2) ;
            \vertex at (-0.5, 0) (b); \vertex[below left=0.0cm of b, black,dot] (b){};
            \vertex at (0.5, 0) (c);  \vertex[below left=0.0cm of c, black,dot] (c){};
            \vertex at (0.125, -0.35) (cc);
            \vertex at (-1, -1) (d); \vertex[below left=0.0cm of d, black, dot] (d){};
            \vertex at (-0.1, -0.5) (dd);
            \vertex at (1, -1) (e); \vertex[below left=0.0cm of e, black, dot] (e){};
            \vertex at (-1,-2) (f) {\(p_1\)};
            \vertex at (1, -2) (g) {\(p_2\)};
            \diagram*{
                (a2) --[plain,thick] (a1),
                (b) -- [plain] (a2),
                (a2) -- [plain] (c),
                (a2) -- [plain] (c),
                (e)-- [plain] (b) ,
                (d) -- [plain] (dd),
                (cc) -- [plain] (c),
                (c) -- [plain] (e),
                (d) -- [plain] (b),
                (f) -- [plain,thick] (d),
                (g) -- [plain,thick] (e),
            };
            \vertex [left=0.4em of a2] {\(3\)};
            \vertex [left=0.6em of b] {\(5\)};
            \vertex [right=0.6em of c] {\(4\)};
            \vertex [left=0.6em of d] {\(2\)};
            \vertex [right=0.6em of e] {\(1\)};
            \node[below=1cm of a1, black, dot] (a2);
        \end{feynman}
              $  \hspace{2cm}   \begin{aligned}
            &=\lim_{p_1,p_2\rightarrow 0}-\frac{i}{12}g^5(C_A- C_A) C_A  \int \prod_{l=1}^5 d^8 z_l~\epsilon_{ijk} [\Phi_{i}(z_1)[\Phi_{j}(z_2), \Phi_{k}(z_3)]]\times \nonumber\\
            &\times\left\{ \frac{1}{\square_1}\delta_{1,3} \frac{D^2_2 \bar{D}^2_3}{16\square_2} \delta_{3,2}   \frac{1}{16 \square_2} \delta_{2,4} \frac{D^2_1\bar{D}^2_4}{16\square_1} \delta_{1,4} \frac{D^2_1\bar{D}^2_5}{16\square_1} \delta_{1,5}\frac{D^2_2}{4\square_2}\delta_{2,5} \right\}=0.
        \end{aligned}
        $
\end{tikzpicture}
        \caption{Feynman two-loop chiral supergraph contributing to the non-holomorphic superpotential.  }
        \label{two-loopChiral}
\end{figure}

In this section, we are going to consider two-loop quantum corrections to the chiral potential.

In the absence of gauge interaction, we have non-planar two-loop diagram which  can explicitly be written as in Fig.\ref{two-loopChiral},
where $\zeta(n)$ is the Riemann zeta-function. This contribution is finite but suppressed by the number of colors $N$, as the corresponding diagram is non-planar. This two-loop finite contribution is purely chiral and contains only chiral interactions. This diagram was first calculated in component form in Ref. \cite{JackJW:1990pd} and in superfield form in Ref. \cite{BKP:1994xq,BKY:1994iw}.
Calculation is consistent with Ref.\cite{Mauri:2006uw}.

\begin{figure}[htbp]
\begin{tikzpicture}[
        scale=1,
        mydot/.style={
            circle,
            fill=white,
            draw,
            outer sep=0pt,
            inner sep=1.5pt
        }
        ]
        \begin{feynman}
            \vertex  at (0,2) (a1){\(p_1+p_2\)};
            \vertex at (0,1) (a2) ;
            \vertex at (-0.5, 0) (b); \vertex[below left=0.0cm of b, black, dot] (b){};
            \vertex at (0.5, 0) (c);  \vertex[below left=0.0cm of c,black, dot] (c){};
            \vertex at (-1, -1) (d); \vertex[below left=0.0cm of d, black, dot] (d){};
            \vertex at (1, -1) (e); \vertex[below left=0.0cm of e, black, dot] (e){};
            \vertex at (-1,-2) (f) {\(p_1\)};
            \vertex at (1, -2) (g) {\(p_2\)};
            \diagram*{
                (a2) --[plain,thick] (a1),
                (b) -- [plain] (a2),
                (a2) -- [plain] (c),
                (a2) -- [plain] (c),
                (c)-- [boson] (b) ,
                (d) -- [plain] (b),
                (c) -- [plain] (e),
                (d) -- [plain] (b),
                (e) -- [boson] (d) ,
                (f) -- [plain,thick] (d),
                (g) -- [plain,thick] (e),
            };
            \vertex [left=0.4em of a2] {\(3\)};
            \vertex [left=0.6em of b] {\(5\)};
            \vertex [right=0.6em of c] {\(4\)};
            \vertex [left=0.6em of d] {\(2\)};
            \vertex [right=0.6em of e] {\(1\)};
            \node[below=1cm of a1, black, dot] (a2);
        \end{feynman}
   $  \hspace{2cm}   \begin{aligned}
         &W^{(A)}=\lim_{p_1,p_2\rightarrow0}\frac{i}{24}g^5 C_A^2\int \prod_{l=1}^5 d^8 z_l~ \epsilon_{ijk} [\Phi_{i}(z_1)[\Phi_{j}(z_2), \Phi_{k}(z_3)]]\times\\& \times\left\{ \frac{1}{\square_1}\delta_{2,1}  \frac{D^2_1 \bar{D}^2_4}{16\square_4}\delta_{1,4} \frac{D^2_4 \bar D^2_3}{16\square_4} \delta_{4,3}  \frac{D^2_5}{4\square_5} \delta_{3,5}   \frac{\bar{D}^2_5 D^2_2}{16\square_5} \delta_{5,2} \frac{1}{\square_4}\delta_{4,5}  \right\}=\\
         &=\frac{1}{4}6g^4(C_A)^2 \zeta(3) \times W_{tree}.
         \end{aligned}  $  \end{tikzpicture}
  \\
 \begin{tikzpicture}[
        scale=1,
        mydot/.style={
            circle,
            fill=white,
            draw,
            outer sep=0pt,
            inner sep=1.5pt
        }
        ]
        \begin{feynman}
            \vertex  at (0,2) (a1){\(p_1+p_2\)};
            \node[below=1cm of a1, mydot] (a2);
            \vertex at (0,1) (a2) ;
            \vertex at (-0.5, 0) (b); \vertex[below left=0.0cm of b, black, dot] (b){};
            \vertex at (0.5, 0) (c);  \vertex[below left=0.0cm of c, black, dot] (c){};
            \vertex at (0.125, -0.35) (cc);
            \vertex at (-1, -1) (d); \vertex[below left=0.0cm of d, black, dot] (d){};
            \vertex at (-0.1, -0.5) (dd);
            \vertex at (1, -1) (e); \vertex[below left=0.0cm of e, black, dot] (e){};
            \vertex at (-1,-2) (f) {\(p_1\)};
            \vertex at (1, -2) (g) {\(p_2\)};
            \diagram*{
                (a2) --[plain,thick] (a1),
                (b) -- [plain] (a2),
                (a2) -- [plain] (c),
                (a2) -- [plain] (c),
                (e)-- [boson] (b) ,
                (d) -- [boson] (dd),
                (cc) -- [boson] (c),
                (c) -- [plain] (e),
                (d) -- [plain] (b),
                (f) -- [plain,thick] (d),
                (g) -- [plain,thick] (e),
            };
            \vertex [left=0.4em of a2] {\(3\)};
            \vertex [left=0.6em of b] {\(5\)};
            \vertex [right=0.6em of c] {\(4\)};
            \vertex [left=0.6em of d] {\(2\)};
            \vertex [right=0.6em of e] {\(1\)};
            \node[below=1cm of a1, black, dot] (a2);
        \end{feynman}
      $  \hspace{2cm}   \begin{aligned}
         &W^{(B)}=0.
         \end{aligned}  $  \end{tikzpicture}
             \\
\begin{tikzpicture}[
        scale=1,
        mydot/.style={
            circle,
            fill=white,
            draw,
            outer sep=0pt,
            inner sep=1.5pt
        }
        ]
        \begin{feynman}
            \vertex  at (0,2) (a1){\(p_1+p_2\)};
            \node[below=1cm of a1, mydot] (a2);
            \vertex at (0,1) (a2) ;
            \vertex at (0.5, 0) (c);  \vertex[below left=0.0cm of c, black, dot] (c){};
            \vertex at (-1, -1) (d); \vertex[below left=0.0cm of d, black, dot] (d){};
            \vertex at (1, -1) (e); \vertex[below left=0.0cm of e, black, dot] (e){};
            \vertex at (-1,-2) (f) {\(p_1\)};
            \vertex at (1, -2) (g) {\(p_2\)};
            \diagram*{
                (a2) --[plain,thick] (a1),
                (d) -- [plain] (a2),
                (a2) -- [plain] (c),
                (e)-- [boson] (d) ,
                (d) -- [boson] (c),
                (c) -- [plain] (e),
                (f) -- [plain,thick] (d),
                (g) -- [plain,thick] (e),
            };
            \vertex [left=0.4em of a2] {\(3\)};
            \vertex [right=0.6em of c] {\(4\)};
            \vertex [left=0.6em of d] {\(2\)};
            \vertex [right=0.6em of e] {\(1\)};
            \node[below=1cm of a1, black, dot] (a2);
        \end{feynman}
       $  \hspace{2cm}   \begin{aligned}
         &W^{(C)}=\lim_{p_1,p_2\rightarrow0}-\frac{i}{24}g^5C_A^2\int \prod_{l=1}^5 d^8 z_l  ~ \epsilon_{ijk} [\Phi_{i}(z_1)[\Phi_{j}(z_2), \Phi_{k}(z_3)]]\times\\& \times \left\{ \frac{1}{\square_1}\delta_{2,1} \frac{D^2_1 \bar{D}^2_4}{16\square_4} \delta_{1,4}   \frac{\bar{D}^2_4 D^2_3}{16\square_4} \delta_{4,3} \frac{D^2_2}{4\square_3} \delta_{3,2}\frac{1}{\square_4}\delta_{2,4} \right\}=\\
         &=\frac{6}{4}g^4C_A^2\zeta(3) \times W_{tree}.
                  \end{aligned}  $  \end{tikzpicture}
             \\
\begin{tikzpicture}[
        scale=1,
        mydot/.style={
            circle,
            fill=white,
            draw,
            outer sep=0pt,
            inner sep=1.5pt
        }
        ]
        \begin{feynman}
            \vertex  at (0,2) (a1){\(p_1+p_2\)};
            \node[below=1cm of a1, mydot] (a2);
            \vertex at (0,1) (a2) ;
            \vertex at (-0.5, 0) (b); \vertex[below left=0.0cm of b, black, dot] (b){};
            \vertex at (0, -1) (c);  \vertex[below left=0.0cm of c,black, dot] (c){};
            \vertex at (-1, -1) (d); \vertex[below left=0.0cm of d, black, dot] (d){};
            \vertex at (1, -1) (e); \vertex[below left=0.0cm of e, black, dot] (e){};
            \vertex at (-1,-2) (f) {\(p_1\)};
            \vertex at (1, -2) (g) {\(p_2\)};
            \diagram*{
                (a2) --[plain,thick] (a1),
                (b) -- [plain] (a2),
                (a2) -- [plain] (e),
                (c) -- [boson] (b),
                (e) -- [boson] (c),
                (c)-- [boson] (d) ,
                (d) -- [plain] (b),
                (f) -- [plain,thick] (d),
                (g) -- [plain,thick] (e),
            };
            \vertex [left=0.4em of a2] {\(3\)};
            \vertex [left=0.6em of b] {\(4\)};
            \vertex [below=0.6em of c] {\(5\)};
            \vertex [left=0.6em of d] {\(2\)};
            \vertex [right=0.6em of e] {\(1\)};
            \node[below=1cm of a1, black, dot] (a2);
        \end{feynman}
       $  \hspace{2cm}   \begin{aligned}
         &W^{(D)}=\lim_{p_1,p_2\rightarrow0}\frac{i}{12}g^5 C_A^2 \int \prod_{l=1}^5 d^8 z_l ~ \epsilon_{ijk} [\Phi_{i}(z_1)[\Phi_{j}(z_2), \Phi_{k}(z_3)]]\times\\& \times \left\{ \frac{1}{\square_5}\delta_{5,1} \frac{D^2_1 \bar{D}^2_3}{16\square_1} \delta_{1,3}  \frac{D^2_4}{4\square_3}\delta_{3,4} \frac{\bar{D}^2_4 D^2_2}{16\square_4} \delta_{4,2}\frac{1}{\square_4}\delta_{2,5} \frac{1}{\square_4}\delta_{5,4}\right\}=\\
         &=\frac{1}{2}g^4 C_A^2 \Upsilon^{(2)} \times W_{tree}.
                  \end{aligned}  $  \end{tikzpicture}
                  \caption{Feynman two-loop finite diagrams that can contribute to the chiral superpotential.  }
        \label{two-loop}
\end{figure}

             \begin{figure}[htbp]
\begin{tikzpicture}[
        scale=1,
        mydot/.style={
            circle,
            fill=white,
            draw,
            outer sep=0pt,
            inner sep=1.5pt
        }
        ]
        \begin{feynman}
            \vertex  at (0,2) (a1){\(p_1+p_2\)};
            \node[below=1cm of a1, mydot] (a2);
            \vertex at (0,1) (a2) ;
            \vertex at (0, -1) (b); \vertex[below left=0.0cm of b, black, dot] (b){};
            \vertex at (0.5, 0) (c);  \vertex[below left=0.0cm of c, black, dot] (c){};
            \vertex at (-1, -1) (d); \vertex[below left=0.0cm of d, black, dot] (d){};
            \vertex at (1, -1) (e); \vertex[below left=0.0cm of e,black, dot] (e){};
            \vertex at (-1,-2) (f) {\(p_1\)};
            \vertex at (1, -2) (g) {\(p_2\)};
            \diagram*{
                (a2) --[plain,thick] (a1),
                (d) -- [plain] (a2),
                (a2) -- [plain] (c),
                (c) -- [boson] (e),
                (e) -- [plain] (b),
                (b)-- [boson] (d) ,
                (c) -- [plain] (b),
                (f) -- [plain,thick] (d),
                (g) -- [plain,thick] (e),
            };
            \vertex [left=0.4em of a2] {\(3\)};
            \vertex [below=0.6em of b] {\(5\)};
            \vertex [right=0.6em of c] {\(4\)};
            \vertex [left=0.6em of d] {\(2\)};
            \vertex [right=0.6em of e] {\(1\)};
            \node[below=1cm of a1, black, dot] (a2);
        \end{feynman}
    $  \hspace{2cm}   \begin{aligned}
         &W^{(E)}=\lim_{p_1,p_2\rightarrow0}\frac{i}{12}g^5C_A^2\int \prod_{l=1}^5 d^8 z_l ~ \epsilon_{ijk} [\Phi_{i}(z_1)[\Phi_{j}(z_2), \Phi_{k}(z_3)]] \times\\& \times\left\{ \frac{1}{\square_4}\delta_{1,4} \frac{D^2_1 \bar{D}^2_5}{16\square_5} \delta_{1,5}\frac{D^2_5 \bar{D}^2_4}{16\square_5} \delta_{5,4} \frac{D^2_4 \bar{D}^2_3}{16\square_4} \delta_{4,3} \frac{D^2_2}{4\square_3} \delta_{3,2}\frac{1}{\square_5}\delta_{2,5} \right\}=\\
         &=-\frac{1}{2} 6g^4(C_A)^2\zeta(3) \times W_{tree}.
                  \end{aligned}  $  \end{tikzpicture}
                \\
\begin{tikzpicture}[
scale=1,
mydot/.style={
    circle,
    fill=white,
    draw,
    outer sep=0pt,
    inner sep=1.5pt
}
]
\begin{feynman}
    \vertex  at (0,2) (a1){\(p_1+p_2\)};
    \node[below=1cm of a1, mydot] (a2);
    \vertex at (0,1) (a2) ;
    \vertex at (-0.5, 0) (b); \vertex[below left=0.0cm of b, black, dot] (b){};
    \vertex at (0.5, 0) (c);  \vertex[below left=0.0cm of c,black, dot] (c){};
    \vertex at (0.125, -0.35) (cc);
    \vertex at (-1, -1) (d); \vertex[below left=0.0cm of d, black, dot] (d){};
    \vertex at (-0.1, -0.5) (dd);
    \vertex at (1, -1) (e); \vertex[below left=0.0cm of e, black, dot] (e){};
    \vertex at (-1,-2) (f) {\(p_1\)};
    \vertex at (1, -2) (g) {\(p_2\)};
    \diagram*{
        (a2) --[plain,thick] (a1),
        (b) -- [plain] (a2),
        (a2) -- [plain] (c),
        (a2) -- [plain] (c),
        (e)-- [boson] (b) ,
        (d) -- [plain] (dd),
        (cc) -- [plain] (c),
        (c) -- [plain] (e),
        (d) -- [plain] (b),
        (f) -- [plain,thick] (d),
        (g) -- [plain,thick] (e),
    };
    \vertex [left=0.4em of a2] {\(3\)};
    \vertex [left=0.6em of b] {\(5\)};
    \vertex [right=0.6em of c] {\(4\)};
    \vertex [left=0.6em of d] {\(2\)};
    \vertex [right=0.6em of e] {\(1\)};
    \node[below=1cm of a1, black, dot] (a2);
\end{feynman}
  $  \hspace{2cm}   \begin{aligned}
         &W^{(F)}\sim\lim_{p_1,p_2\rightarrow0} 4g^4\int \prod_{l=1}^5 d^8 z_l   \epsilon_{ijk}[\Phi_{i}(z_1)[\Phi_{j}(z_2), \Phi_{k}(z_3)]] \times\\& \times\left\{ \frac{1}{\square_4}\delta_{2,4} \frac{D^2_1 \bar{D}^2_4}{16\square_4} \delta_{1,4} \frac{\bar{D}^2_4 D^2_3}{16\square_4} \delta_{4,3} \frac{D^2_5}{4\square_5} \delta_{3,5} \frac{\bar{D}^2_5 D^2_2}{16\square_5} \delta_{5,2}\frac{1}{\square_5}\delta_{1,5} \right\}=0.
                  \end{aligned}  $  \end{tikzpicture}
            \\
\begin{tikzpicture}[
scale=1,
mydot/.style={
    circle,
    fill=white,
    draw,
    outer sep=0pt,
    inner sep=1.5pt
}
]
\begin{feynman}
    \vertex  at (0,2) (a1){\(p_1+p_2\)};
    \node[below=1cm of a1, red, dot] (a2);
    \vertex  at (0,0) (a3); \vertex[below left=0.0cm of a3, empty dot] (a3){};
    \vertex at (0,1) (a2) ;
    \vertex at (-1, -1) (b); \vertex[below left=0.0cm of b, red,dot] (b){};
    \vertex at (1, -1) (c);  \vertex[below left=0.0cm of c,red, dot] (c){};
    \vertex at (-2,-2) (f) {\(p_1\)};
    \vertex at (2, -2) (g) {\(p_2\)};
    \diagram*{
        (a2) --[plain,thick] (a1),
        (a2) --[plain] (a3),
        (b) -- [plain] (a3),
        (a3) -- [plain] (c),
        (a2) -- [boson] (b),
        (a2)-- [boson] (c) ,
        (f) -- [plain,thick] (b),
        (g) -- [plain,thick] (c),
    };
    \vertex [left=0.4em of a2] {\(3\)};
    \vertex [left=0.6em of b] {\(2\)};
    \vertex [right=0.6em of c] {\(1\)};
    \vertex [right=0.6em of a3] {\(4\)};
    \vertex[below left=0.0cm of d, black, dot] (d){};
    \vertex[below left=0.0cm of b, black, dot] (b){};
    \vertex[below left=0.0cm of c, black, dot] (c){};
    \node[below=1cm of a1, black, dot] (a2);
    \vertex[below left=0.0cm of a3, black, dot] (a3){};
\end{feynman}
     $  \hspace{2cm}   \begin{aligned}
         &W^{(G)} \sim \lim_{p_1,p_2\rightarrow0}4g^4\int \prod_{l=1}^5 d^8 z_l  \epsilon_{ijk} [\Phi_{i}(z_1)[\Phi_{j}(z_2), \Phi_{k}(z_3)]]\times\\& \times\left\{ \frac{D^2_1 \bar{D}^2_4}{16\square_4}\delta_{1,4} \frac{\bar{D}^2_4 D^2_2}{16\square_4} \delta_{4,2}  \frac{1}{\square_3} \delta_{2,3} \frac{1}{\square_1} \delta_{3,1} \frac{D^2_3}{4\square_4} \delta_{3,4} \right\}=0.
                  \end{aligned}  $  \end{tikzpicture}
               \\
                \begin{tikzpicture}[
scale=1,
mydot/.style={
    circle,
    fill=white,
    draw,
    outer sep=0pt,
    inner sep=1.5pt
}
]
\begin{feynman}
    \vertex  at (0,2) (a1){\(p_1+p_2\)};
    \node[below=1cm of a1, red, dot] (a2);
    \vertex  at (0,0) (a3); \vertex[below left=0.0cm of a3, empty dot] (a3){};
    \vertex at (0,1) (a2) ;
    \vertex at (-1, -1) (b); \vertex[below left=0.0cm of b, red,dot] (b){};
    \vertex at (1, -1) (c);  \vertex[below left=0.0cm of c,red, dot] (c){};
    \vertex at (0, -1) (d); \vertex[below left=0.0cm of d, red,dot] (d){};
    \vertex at (-2,-2) (f) {\(p_1\)};
    \vertex at (2, -2) (g) {\(p_2\)};
    \diagram*{
        (a2) --[plain,thick] (a1),
        (a2) --[plain] (a3),
        (b) -- [plain] (a3),
        (a3) -- [plain] (c),
        (b) -- [boson] (c),
        (a2)-- [boson, quarter right, looseness=1.5] (d) ,
        (f) -- [plain,thick] (b),
        (g) -- [plain,thick] (c),
    };
    \vertex [left=0.4em of a2] {\(3\)};
    \vertex [left=0.6em of b] {\(2\)};
    \vertex [right=0.6em of c] {\(1\)};
    \vertex [right=0.6em of a3] {\(4\)};
    \vertex[below left=0.0cm of d, black,dot] (d){};
    \vertex[below left=0.0cm of b, black,dot] (b){};
    \vertex[below left=0.0cm of c, black,dot] (c){};
    \node[below=1cm of a1, black, dot] (a2);
    \vertex[below left=0.0cm of a3, black, dot] (a3){};
\end{feynman}
             $  \hspace{2cm}   \begin{aligned}
         &W^{(H)}\sim\lim_{p_1,p_2\rightarrow0}2g^4\int \prod_{l=1}^5 d^8 z_l ~\epsilon_{ijk} [\Phi_{i}(z_1)[\Phi_{j}(z_2), \Phi_{k}(z_3)]]\times\\& \times\left\{ \frac{1}{\square_5} \delta_{3,5}\frac{D^2_1 \bar{D}^2_4}{16\square_4}\delta_{1,4} \frac{\bar{D}^2_4 D^2_2}{16\square_4} \delta_{4,2}  \frac{1}{\square_3} \delta_{2,5} \frac{1}{\square_1} \delta_{5,3} \frac{D^2_3}{4\square_4} \delta_{3,4} \right\}=0.
                  \end{aligned}  $  \end{tikzpicture}
        \caption{Feynman two-loop finite diagrams that can contribute to the chiral superpotential.  }
        \label{two-loop2}
\end{figure}

We also compute other finite Feynman supergraphs, see Figs.~\ref{two-loop}-\ref{two-loop2}.
Diagram $W^{(F)}$ after  $D$-algebra transformations produces the $\bar{D}^2\Phi$ term which is identically zero.
The last two master-integrals  $W^{(G,H)}$ are equal to zero due to the adopted Fermi-Feynman gauge.

So finally restoring all factors of the $SU(N)$ gauge group, we have the following result for the two-loop finite chiral effective superpotential
\begin{equation}
    \textbf{W}^{(2)}_{fin} =C_A^2\left[g^4\Upsilon^{(2)} \right] \times W_{tree},
    \label{finWN1}
\end{equation}
where $\Upsilon^{(2)}$ is the two-loop Usyukina-Davydychev triangle function \cite{Usyukina:1992jd}
\begin{align}
\Upsilon^{(2)}=\int_0^1 d\tau \frac{2 \log^3(\tau)}{\tau^2-\tau+1}.
\end{align}

\section{Ladder limit}\label{sec:largeN}

\begin{figure}[htbp]
\centering
 \begin{minipage}[h]{0.2\linewidth}
    \vspace{0pt}
      \begin{subfigure}{.3\textwidth}
        \centering
    \begin{tikzpicture}[
        scale=1,
        mydot/.style={
            circle,
            fill=white,
            draw,
            outer sep=0pt,
            inner sep=1.5pt
        }
        ]
        \begin{feynman}
            \vertex  at (0,2) (a1){\(p_1+p_2\)};
            \node[below=1cm of a1, mydot] (a2);
            \vertex at (0,1) (a2) ;
            \vertex at (-0.5,0) (q); 
            \vertex at (0, -1) (w);  
            \vertex at (-0.75, -0.5) (b); \vertex[below left=0.0cm of b, black, dot] (b){};
            \vertex at (-0.25, 0.5) (bb);
            \vertex at (-0.5, -1) (c);  \vertex[below left=0.0cm of c,black, dot] (c){};
            \vertex at (0.5, -1) (cc);
            \vertex at (-1, -1) (d); \vertex[below left=0.0cm of d, black, dot] (d){};
            \vertex at (1, -1) (e); \vertex[below left=0.0cm of e, black, dot] (e){};
            \vertex at (-1,-2) (f) {\(p_1\)};
            \vertex at (1, -2) (g) {\(p_2\)};
            \diagram*{
                (a2) --[plain,thick] (a1),
                (b) -- [plain] (a2),
                (a2) -- [plain] (e),
                (c) -- [boson] (b),
                (cc) -- [boson] (bb),
                (e) -- [boson] (c),
                (c)-- [boson] (d) ,
                (d) -- [plain] (b),
                (f) -- [plain,thick] (d),
                (g) -- [plain,thick] (e),
            };
            \vertex [left=0.4em of a2] {\(3\)};
               \vertex [right=1.2em of b] {\(\ldots\)};
        \vertex [left=0.6em of d] {\(2\)};
        \vertex [right=0.6em of e] {\(1\)};
                \node[below=1cm of a1,black, dot] (a2);
        \vertex[below left=0.0cm of cc,black, dot] (cc){};
        \vertex[below left=0.0cm of bb, black, dot] (bb){};
        \end{feynman}
    \end{tikzpicture}   \label{fig:a}\caption{}
   \end{subfigure}
       \end{minipage}
         \begin{minipage}[h]{0.2\linewidth}
    \vspace{0pt}
      \begin{subfigure}{.3\textwidth}
        \centering
        \begin{tikzpicture}[
            scale=1,
            mydot/.style={
                circle,
                fill=white,
                draw,
                outer sep=0pt,
                inner sep=1.5pt
            }
            ]
            \begin{feynman}
                \vertex  at (0,2) (a1){\(p_1+p_2\)};
                \node[below=1cm of a1, black, dot] (a2);
                \vertex at (0,1) (a2) ;
                \vertex at (-0.5, 0) (b); 
                \vertex at (0.5, 0) (c);  
                \vertex at (-0.25, 0.5) (bb);
                \vertex at (0.25, 0.5) (cc);
                \vertex at (-0.75, -0.5) (bbb);
                \vertex at (0.75, -0.5) (ccc);
                \vertex at (-1, -1) (d); \vertex[below left=0.0cm of d, black, dot] (d){};
                \vertex at (1, -1) (e); \vertex[below left=0.0cm of e, black, dot] (e){};
                \vertex at (-1,-2) (f) {\(p_1\)};
                \vertex at (1, -2) (g) {\(p_2\)};
                \diagram*{
                    (a2) --[plain,thick] (a1),
                    (b) -- [plain] (a2),
                    (a2) -- [plain] (c),
                    (cc)-- [boson] (bb) ,
                    (ccc)-- [boson] (bbb) ,
                    (d) -- [plain] (b),
                    (c) -- [plain] (e),
                    (d) -- [plain] (b),
                    (e) -- [boson] (d) ,
                    (f) -- [plain,thick] (d),
                    (g) -- [plain,thick] (e),
                };
                \vertex [left=0.4em of c] {\(\ldots\)};
                \node[below=1cm of a1, black, dot] (a2);
                \vertex[below left=0.0cm of bb, black, dot] (bb){};
                \vertex[below left=0.0cm of cc,black, dot] (cc){};
                \vertex[below left=0.0cm of bbb, black, dot] (bbb){};
                \vertex[below left=0.0cm of ccc,black, dot] (ccc){};
            \end{feynman}
        \end{tikzpicture} \label{fig:b}\caption{}
      \end{subfigure}
       \end{minipage}
                \begin{minipage}[h]{0.2\linewidth}
    \vspace{0pt}
        \begin{subfigure}{.3\textwidth}
            \centering
            \begin{tikzpicture}[
                scale=1,
                mydot/.style={
                    circle,
                    fill=white,
                    draw,
                    outer sep=0pt,
                    inner sep=1.5pt
                }
                ]
                \begin{feynman}
                    \vertex  at (0,2) (a1){\(p_1+p_2\)};
                    \node[below=1cm of a1, mydot] (a2);
                    \vertex at (0,1) (a2) ;
                    \vertex at (0.75, -0.5) (b); \vertex[below left=0.0cm of b, black, dot] (b){};
                    \vertex at (0.25, 0.5) (bb);
                    \vertex at (-1, -1) (d); \vertex[below left=0.0cm of d, black, dot] (d){};
                    \vertex at (1, -1) (e); \vertex[below left=0.0cm of e, black, dot] (e){};
                    \vertex at (-1,-2) (f) {\(p_1\)};
                    \vertex at (1, -2) (g) {\(p_2\)};
                    \diagram*{
                        (a2) --[plain,thick] (a1),
                        (d) -- [plain] (a2),
                        (a2) -- [plain] (e),
                        (e)-- [boson] (d) ,
                        (d) -- [boson] (b),
                        (d) -- [boson] (bb),
                        (f) -- [plain,thick] (d),
                        (g) -- [plain,thick] (e),
                    };
                    \vertex [left=0.4em of a2] {\(3\)};
                    \vertex [below=1.5em of bb] {\(\ldots\)};
                    \vertex [left=0.6em of d] {\(2\)};
                    \vertex [right=0.6em of e] {\(1\)};
                    \node[below=1cm of a1, black, dot] (a2);
                    \vertex[below left=0.0cm of bb, black, dot] (bb){};
                     \vertex[below left=0.0cm of b, black, dot] (b){};
                \end{feynman}
            \end{tikzpicture} \label{fig:c}\caption{}
        \end{subfigure}
         \end{minipage}
\caption{Series of ladder type diagrams.}
\label{fig:ladder1}
\end{figure}
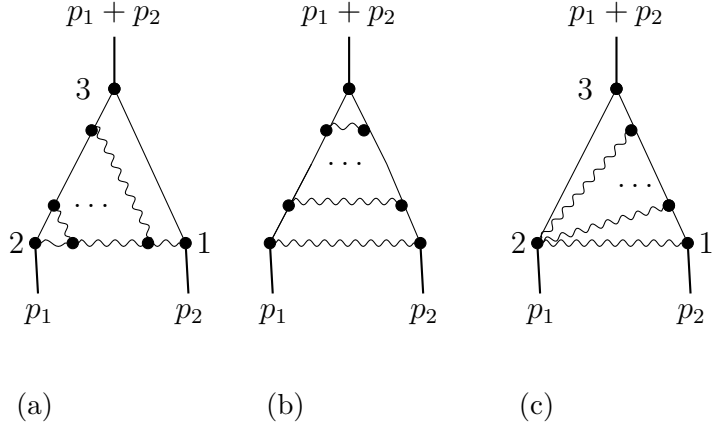

In this section we consider a series of ladder triangle diagrams (see, e.g. Fig.\ref{fig:ladder1})  which can be easily calculated in the $SU(N)$ gauge $\mathcal{N}=4$ super-Yang-Mills theory.
It turns out that we can sum the full set of triangle-type diagrams and reduce them to the sequence of scalar master diagrams. We manually checked these diagrams from the first two-loops described in the previous section up to the fifth loop order utilizing \texttt{SusyMath.m}\cite{Ferrari:2007sc} and found their predictive structure. In the following we discuss these diagrams in more detail.

Let us consider the diagram depicted in Fig. \ref{fig:ladder1} (a):
\begin{align}
    \textbf{W}'^{(m)} &=\lim_{p_1,p_2\rightarrow0}(-1)^{m+1}\frac{g^{2m}}{24} C_A^{m} \int \prod_{l=1}^{2m+1} d^8 z_l ~ g \epsilon_{ijk} [\Phi_i(z_1)[\Phi_j(z_2),\Phi_k(z_3)]] \times\nonumber \\ \times & \left\{ \frac{1}{\square_{5}}\delta_{5,1} \frac{D^2_1 \bar{D}^2_3}{16\square_1} \delta_{1,3} \frac{D^2_{4}}{4\square_3}\delta_{3,4} \frac{\bar{D}^2_4 D^2_6}{16\square_4}  \delta_{4,6}\ldots \right. \nonumber\\& \left.  \dots\frac{\bar{D}_{2m-2}^2 D_{2m}^2}{\square_{2m-2}}\delta_{2m-2,2m}\frac{\bar{D}^2_{2m}D^2_2}{\square_{2m}}\delta_{2m,2}\frac{1}{\square_{2m+1}}\delta_{2m+1,2}\ldots \frac{1}{\square_{2m}}\delta_{2m+1,2m} \ldots \frac{1}{\square_4}\delta_{5,4}\right\}=\nonumber\\
    &=(-1)^{m-1}\frac{g^{2m}}{4} C_A^{m}\Upsilon^{(m)} \times W_{tree}.
\end{align}
Here we first remove all free delta functions over Grassmann variables and use the anticommutation relations for covariant derivatives (see Ref. \cite{BK:book}).
Then evaluate $D$-algebra and re-express the integral through the $\Upsilon$-functions \footnote{
$\Upsilon^{(l)}=\int_0^1 d\tau \frac{2 \log^{2l-1}(\tau)}{\tau^2-\tau+1}=\frac{1}{2^{2l-1} 3^{2l}}  \left(\psi ^{(2l+1)}\left(\frac{2}{3}\right)-\psi ^{(2l+1)}\left(\frac{1}{3}\right)-\psi ^{(2l+1)}\left(\frac{1}{6}\right)+\psi ^{(2l+1)}\left(\frac{5}{6}\right)\right)$, where $\psi^{(n)}$ is the polygamma function. Further simplifications of these functions can be carried out in terms of the multiple zeta or Clausen functions.}. This expression generalizes the finite contributions found in the previous section.
The formal sums $\textbf{W}'^{lead}=\sum \textbf{W}'^{(m)} $ over these contributions (assuming $y \log (\tau)\leq1$ which is justified in the large $N$ limit where $y=g^2 C_A$) are given in compact form after taking the limit $p_{1},p_2\rightarrow0$ in the following expression:
\begin{equation}
\textbf{W}'^{lead}=\frac{g}{2}\int_0^1~d \tau \frac{\log(\tau)~(1-\tau)}{(1+y \log^2(\tau))~(1+\tau^3) } \times W_{tree}=\Upsilon^{tot} \times W_{tree},
\end{equation}
and after integration one has the closed result
\begin{equation}
    \Upsilon^{tot}=\frac{g}{4}\sum_{m=1}^\infty ((\pi -2 \text{Si}(x)) \sin (x)-2 \text{Ci}(x) \cos (x))U_m\left(1/2\right),
\end{equation}
where $x=\frac{m+1}{\sqrt{y}}$ and $\text{Si}$ and $\text{Ci}$ are the integral sine and cosine functions and $U_{n}(x)$ are Chebyshev polynomials.  That is, this expression exactly sums up all the leading color contributions to the chiral effective superpotential of the finite $\mathcal{N}=4$ SYM theory.
Note that  $ \Upsilon^{tot}\to 0$ at $\bar{h} \to 0$. It would be interesting to establish a connection between such contributions and the dual string theory.

It is quite possible that there is a hidden symmetry allowing to
find contributions for all higher loop superdiagrams in the
$\mathcal{N}=4$ SYM and its $\beta$-deformed version by perturbation
theory and find a closed-form expression for the whole series at
least under some assumptions or approximations, we leave the study
of this problems for future works.

\section{Discussion}\label{sec:Concl}

We have performed  calculations of the chiral effective potential in
the $\mathcal{N}=4$ supersymmetric gauge theory formulated in terms
of $\mathcal{N}=1$ superfields. In such a  formulation, the theory
is characterized by chiral superpotential to which quantum
corrections are computed. Note that in the theory under
consideration, all the supergraphs, that give rise to the chiral
effective potential, are automatically finite and we do not need to
carry out a renormalization of the subgraphs.

The computation of the loop contributions to chiral effective
potential was given using the Wolfram Mathematica packages (e.g.,
\texttt{SusyMath.m} \cite{Ferrari:2007sc}) and methods for computing
multi-loop scalar master-integrals. In the considered finite
$\mathcal{N}=4$ super Yang-Mills model, we obtained the exact sum of
all contributions to the chiral effective potential and found the
result exact in coupling constant. In addition summation of specific
individual sequences of diagrams have been shown in the large $N$
limit. The final result for chiral effective potential is extremely
simple and clear: the effective potential is proportional to the
classical superpotential, where all the details of quantum theory
are concentrated in a single proportionality coefficient.

\subsection*{Acknowledgements}
The authors are grateful to A. A. Tseytlin for interest to work, useful comments and for independent check of the relation $\lim_{\bar h\to \infty} \Upsilon^{tot}=0$ and  P. West for correspondence.
A. Mukhaeva's work is supported by the Foundation for the
Advancement of Theoretical Physics and Mathematics BASIS, No
24-1-4-36-1.

\subsection*{Conflict of Interest} The authors declare that they have no conflicts of interest

\bibliographystyle{JHEP}
\bibliography{refs}
\end{document}